\documentclass[12pt]{article}
\pdfoutput=1

\usepackage{epsfig}
\usepackage{psfrag}
\usepackage{latexsym}
\usepackage{indentfirst}
\usepackage{fancyhdr}
\usepackage{dsfont}
\usepackage{amssymb}
\usepackage{amsmath}
\usepackage{amsfonts}
\usepackage{pifont}
\usepackage{cite}
\usepackage{bbold}
\usepackage{color}
\usepackage{colordvi}
\usepackage{fancybox}
\usepackage[colorlinks]{hyperref}
\usepackage[footnotesize]{caption2}
\usepackage{graphicx}
\usepackage[center,footnotesize,hang]{subfigure}
\usepackage{bbm}
\usepackage{url}
\usepackage{longtable}
\usepackage{multirow}
\usepackage{array}
\usepackage{arydshln}
\usepackage{ulem}
\usepackage{enumitem}
\newcommand{\PreserveBackslash}[1]{\let\temp=\\#1\let\\=\temp}
\newcolumntype{C}[1]{>{\PreserveBackslash\centering}p{#1}}
\newcolumntype{R}[1]{>{\PreserveBackslash\raggedleft}p{#1}}
\newcolumntype{L}[1]{>{\PreserveBackslash\raggedright}p{#1}}
\allowdisplaybreaks

\newcommand{\bq}{\begin{eqnarray}}
\newcommand{\nq}{\end{eqnarray}}

\def\bvec#1{\raise1.5ex\hbox{$\rightarrow$}\mkern-16.5mu #1}

\begin{document}

\title{\hfill ~\\[-30mm] \hfill\mbox{\small USTC-ICTS-16-10}\\[10mm]
              \textbf{$A_4$ and CP symmetry and a model with maximal CP  violation}}

\date{}

\author{\\[1mm]Cai-Chang Li\footnote{E-mail: {\tt lcc0915@mail.ustc.edu.cn}}~,~~Jun-Nan Lu\footnote{E-mail: {\tt hitman@mail.ustc.edu.cn}}~,~~Gui-Jun Ding\footnote{E-mail: {\tt dinggj@ustc.edu.cn}}\\ \\
\it{\small Interdisciplinary Center for Theoretical Study and  Department of Modern Physics, }\\
  \it{\small University of Science and
    Technology of China, Hefei, Anhui 230026, China}\\[4mm]
     }
\maketitle

\begin{abstract}
We study a second CP symmetry compatible with the $A_4$ flavor group, which interchanges the representations $\mathbf{1}'$ and $\mathbf{1}''$. We analyze the lepton mixing patterns arising from the $A_4$ and CP symmetry broken to residual subgroups $Z_3$ and $Z_2\times CP$ in the charged lepton and neutrino sectors respectively. One phenomenologically viable mixing pattern is found, and it predicts maximal atmospheric mixing angle as well as maximal Dirac CP phase, trivial Majorana phase and the correlation $\sin^2\theta_{12}\cos^2\theta_{13}=1/3$. We construct a concrete model based on the $A_4$ and CP symmetry, the above interesting mixing pattern is achieved, the observed charged lepton mass hierarchy is reproduced, and the reactor mixing angle $\theta_{13}$ is of the correct order.

\end{abstract}
\thispagestyle{empty}
\vfill

\newpage
\setcounter{page}{1}

\section{Introduction}

The origin of flavor mixing is one of the most fascinating unsolved problems in particle physics. The precise measurement of neutrino oscillation provides us new window to solve the flavor puzzles. The most popular approach to understand the observed lepton mixing pattern is based on the assumption that a flavor symmetry group (usually finite and non-abelian) is broken down to different subgroups in the charged lepton and neutrino sectors and the mismatch between the two subgroups allows one to predict the lepton mixing matrix up to permutations of rows and columns.
A prime example is the famous tri-bimaximal mixing which can
be derived from simple flavor groups such as $A_4$~\cite{A4_works}
and $S_4$~\cite{Ma:2005pd}. There is an extensive literature on study of different flavor symmetries and their application in model building, please see Refs.~\cite{Altarelli:2010gt,Ishimori:2010au,King:2013eh} for review and additional references.

The discovery of a relatively large reactor mixing angle $\theta_{13}\simeq0.15$ measured by Daya Bay~\cite{An:2012eh}, RENO~\cite{Ahn:2012nd} and Double Chooz~\cite{Abe:2011fz} rules out the tri-bimaximal mixing pattern, and it lead us to scrutinize the discrete flavor symmetry approach. For example, comprehensive surveys of finite non-abelian subgroups of SU(3) and U(3) reveal that mixing angles in agreement with experimental data can be obtained from some large flavor symmetry groups (e.g. $(Z_{18}\times Z_6)\rtimes S_3$ with the group id [648, 259]), while the Dirac CP phase is predicted to be trivial~\cite{Holthausen:2012wt,Fonseca:2014koa,Yao:2015dwa}. Another approach is to amend the flavor symmetry with a CP symmetry~\cite{Ecker:1981wv,Grimus:1995zi,Feruglio:2012cw,Holthausen:2012dk}. In this case, the CP symmetry is represented by a CP transformation which acts non-trivially on flavor space, and consequently it is also dubbed as generalized CP symmetry. In order to consistently combine flavor symmetry with CP symmetry, the so-called consistency conditions have to be fulfilled~\cite{Grimus:1995zi,Feruglio:2012cw,Holthausen:2012dk,Chen:2014tpa} and thus the explicit from of the CP transformation is strongly constrained. Similar to the paradigm of flavor symmetry, the lepton mixing matrix is completely fixed by the residual symmetry of the lepton mass matrices~\cite{Chen:2014wxa,Chen:2015nha,Everett:2015oka,Chen:2015siy,Chen:2016ica} and no predictions can be made for the lepton masses. The advantage of CP symmetry over flavor symmetry is that the CP symmetry can constrain the Dirac as well as Majorana CP violating phases~\cite{Chen:2014wxa,Chen:2015nha,Everett:2015oka,Chen:2015siy,Chen:2016ica}, and a non-vanishing reactor mixing angle can be explained from a small flavor symmetry group. There have been many interesting work studying the predictions of models with both flavor and CP symmetries for a variety of groups such as $A_{4}$~\cite{Feruglio:2012cw,Ding:2013bpa,Nishi:2016jqg}, $S_{4}$~\cite{Feruglio:2012cw,Mohapatra:2012tb,Ding:2013hpa,Li:2014eia,Feruglio:2013hia,Luhn:2013vna, Li:2013jya}, $\Delta(27)$~\cite{Branco:2015hea,Branco:2015gna}, $\Delta(48)$~\cite{Ding:2013nsa}, $A_{5}$~\cite{Li:2015jxa,DiIura:2015kfa,Ballett:2015wia,Turner:2015uta}, $\Delta(96)$~\cite{Ding:2014ssa}, $\Sigma(36\times 3)$~\cite{Rong:2016cpk}, SU(3) infinite group series $\Delta(3n^{2})$~\cite{Hagedorn:2014wha,Ding:2015rwa}, $\Delta(6n^{2})$~\cite{Hagedorn:2014wha,King:2014rwa,Ding:2014ora} and $D^{(1)}_{9n, 3n}$~\cite{Li:2016ppt} for a general integer $n$. Recently a comprehensive scan of all finite discrete groups with order less than 2000 is performed, the CP symmetry corresponding to class-inverting automorphism of the flavor group is imposed, and the possible lepton mixing patterns are presented~\cite{Yao:2016zev}. Furthermore, the flavor and CP symmetries can make clear predictions for  neutrinoless double beta~\cite{Ding:2013hpa,Li:2014eia,Li:2015jxa,Ding:2014ora,Li:2016ppt,Hagedorn:2016lva,Yao:2016zev} and leptogenesis~\cite{Hagedorn:2016lva,Chen:2016ptr,Yao:2016zev}. In particular the low energy Dirac and Majorana CP violating phases are connected to the CP violation in leptogenesis in this scenario~\cite{Chen:2016ptr,Yao:2016zev}.

$A_4$ is the minimal group which has three-dimensional irreducible representation. $A_4$ as a flavor symmetry group has been extensively studied,  and a vast number of models are constructed. The interplay between $A_4$ and CP symmetry has been studied as well~\cite{Feruglio:2012cw,Ding:2013bpa,Nishi:2016jqg}. The recent studies involving $A_4$ and CP can be seen in Refs.~\cite{He:2015afa,Ma:2015pma,He:2015gba}. It turns out that two possible CP symmetries can be defined in the context of $A_4$~\cite{Ding:2013bpa}. The first one acts on the fields as
\begin{equation}
\label{eq:CP_first}
\varphi_{\mathbf{3}}\rightarrow \varphi^{*}_{\mathbf{3}},\quad \varphi_{\mathbf{1}'}\rightarrow \varphi^{*}_{\mathbf{1}'},\quad \varphi_{\mathbf{1}''}\rightarrow \varphi^{*}_{\mathbf{1}''}\,,
\end{equation}
in our working basis, where $\varphi_{\mathbf{r}}$ denotes a field transforming as $A_4$ irreducible representation $\mathbf{r}$. This CP symmetry can be imposed in a generic $A_4$ model regardless of the field content. Its predictions for lepton mixing arising from all possible residual symmetries are comprehensively studied in Ref.~\cite{Ding:2013bpa}, and a model is built to realize the model independent predictions. The second possible CP symmetry compatible with $A_4$ is given by~\cite{Ding:2013bpa}
\begin{equation}
\label{eq:CP_second}
\varphi_{\mathbf{3}}\rightarrow\begin{pmatrix}
1 &~  0  &~  0\\
0 &~  0  &~  1 \\
0 &~  1  &~  0
\end{pmatrix}
\varphi^{*}_{\mathbf{3}},\quad \varphi_{\mathbf{1}'}\rightarrow \varphi^{*}_{\mathbf{1}''},\quad \varphi_{\mathbf{1}''}\rightarrow \varphi^{*}_{\mathbf{1}'}\,.
\end{equation}
In this case, both $\varphi_{\mathbf{1}'}$ and $\varphi_{\mathbf{1}''}$ should be present in pair or absent simultaneously and they should carry the same quantum numbers under all the symmetries of the model except $A_4$. In the present work we shall be concerned with this second CP symmetry.

The article is organized as follows. In section~\ref{sec:revisited_A4} we revisit the possible CP symmetries which are compatible with the $A_4$ flavor symmetry, and our conventions for the $A_4$ group and its representations are presented. In section~\ref{sec:general_analysis} we study the lepton mixing patterns which arise from the breaking of the $A_4$ and CP symmetry to the remnant symmetries $Z_3$ in the charged lepton sector and to $Z_2\times CP$ in the neutrino sector. We find one phenomenologically viable case in which both atmospheric mixing angle and Dirac CP phase are maximal and the sum rule $\sin^2\theta_{12}\cos^2\theta_{13}=1/3$ between the solar and the reactor mixing angles is fulfilled. In section~\ref{sec:model} we construct an explicit model based on $A_4$ and CP symmetry, the lepton mixing matrix is of the tri-maximal form at leading order, the subleading corrections generate the correct size of the reactor mixing angle, and the interesting mixing pattern found in section~\ref{sec:general_analysis} is realized exactly. Finally, section~\ref{sec:conclusion} is devoted to our conclusion.

\section{\label{sec:revisited_A4} Revisiting $A_{4}$ and generalized CP symemtry}

$A_{4}$ is the symmetry group of the tetrahedron. It contains twelve
elements and it is the smallest non-abelian finite group which admits an irreducible three-dimensional representation. $A_4$ can be generated by two generators $S$ and $T$ obeying the relations~\cite{Altarelli:2005yx}
\begin{equation}
S^{2}=T^{3}=(ST)^{3}=1\,.
\end{equation}
There are four equivalence classes (two elements $a$ and $b$ belong to the same equivalence class if there exists a group element $g$ such that $gag^{-1}=b$):
\begin{eqnarray}
\nonumber &1C_1=\{1\}\,, \qquad &3C_2=\{S,TST^2,T^2ST\}\,,\\
&4C_3= \{T,ST,TS,STS\}\,, \qquad & 4C^{\prime}_3=\{ T^2,ST^2,T^2S,ST^2S\}\,,
\end{eqnarray}
where $kC_{n}$ denotes a conjugacy class which contains $k$ elements with order $n$. The $A_4$ group has four inequivalent irreducible representations: three singlets $\mathbf{1}$, $\mathbf{1}^{\prime}$, $\mathbf{1}^{\prime\prime}$ and a triplet $\mathbf{3}$. The three one-dimensional representations are given by
\begin{eqnarray}
\nonumber && \mathbf{1}:~~ S=1, \qquad T=1 \,,  \\
\nonumber && \mathbf{1}^{\prime}:~~ S=1, \qquad T=\omega^{2} \,,  \\
 &&\mathbf{1}^{\prime\prime}:~~S=1, \qquad T=\omega \,,
\end{eqnarray}
where $\omega=e^{i2\pi/3}$.  The three-dimensional representation $\mathbf{3}$, in a basis where the generator $T$ is diagonal, is constructed from
\begin{equation}
S=\frac{1}{3}\left(\begin{array}{ccc}
    -1& 2  & 2  \\
    2  & -1  & 2 \\
    2 & 2 & -1
\end{array}\right),  \qquad
T=\left(\begin{array}{ccc}
    1 &~ 0 ~& 0 \\
    0 &~ \omega^{2} ~& 0 \\
    0 &~ 0 ~& \omega
\end{array}\right) \,.
\end{equation}
Thus, from two such triplets $\alpha=(\alpha_1,\alpha_2,\alpha_3)\sim\mathbf{3}$ and  $\beta=(\beta_1,\beta_2,\beta_3)\sim\mathbf{3}$ we can obtain the irreducible representations from their product~\cite{Altarelli:2005yx}:
\begin{eqnarray}
\nonumber &&(\alpha\beta)_{\mathbf{1}}=\alpha_1\beta_1+\alpha_2\beta_3+\alpha_3\beta_2\,, \\
\nonumber &&(\alpha\beta)_{\mathbf{1}^{\prime}}=\alpha_3\beta_3+\alpha_1\beta_2+\alpha_2\beta_1\,, \\
\nonumber &&(\alpha\beta)_{\mathbf{1}^{\prime\prime}}=\alpha_2\beta_2+\alpha_1\beta_3+\alpha_3\beta_1\,, \\
\nonumber &&(\alpha\beta)_{\mathbf{3}_S}=\frac{1}{3}(
2\alpha_1\beta_1-\alpha_2\beta_3-\alpha_3\beta_2 ,
2\alpha_3\beta_3-\alpha_1\beta_2-\alpha_2\beta_1,
2\alpha_2\beta_2-\alpha_1\beta_3-\alpha_3\beta_1),\\
\label{eq:CG_coefficient} &&(\alpha\beta)_{\mathbf{3}_A}=\frac{1}{2}(
\alpha_2\beta_3-\alpha_3\beta_2,
\alpha_1\beta_2-\alpha_2\beta_1,
\alpha_3\beta_1-\alpha_1\beta_3)\,,
\end{eqnarray}
where the subscript $S$ ($A$) denotes symmetric (antisymmetric) combination. In the present work we shall study the popular $A_4$ flavor symmetry combined with the generalized CP symmetry. The action of a generalized CP transformation $X$ on a field multiplet $\varphi(x)$ is
\begin{equation}
\varphi(x)\stackrel{CP}{\longrightarrow}X\,\varphi^{*}(x_{P})\,,
\end{equation}
where $x_{P}=(t,-\vec{x})$ and the obvious action of CP on the spinor indices is omitted for the case of $\varphi$ being spinor. Since the CP transformation $X$ acts nontrivially on the flavor space, the consistency condition between flavor and CP symmetries must be satisfied~\cite{Feruglio:2012cw,Holthausen:2012dk,Chen:2014tpa}
\begin{equation}
\label{eq:consistency}
X\rho^*(g)X^{\dagger}=\rho(g^{\prime}), \quad g,g^{\prime} \in A_{4}\,,
\end{equation}
where $\rho(g)$ is the representation matrix of the group element $g$, and it is generally reducible. To be more specific, $\rho(g)$ is generally the direct sum of the $A_4$ irreducible representations corresponding to the particle content of the model. Obviously the elements $g$ and $g'$ should be of the same order. Moreover, given a viable CP transformation $X$, other new CP transformation of the form $\rho(h)X$ for any $h\in A_4$ can be generated if one first performs a flavor symmetry transformation $\rho(h)$ and subsequently the CP transformation $X$. Accordingly the consistency equation is of the form
\begin{equation}
\left[\rho(h)X\right]\rho^*(g)\left[\rho(h)X\right]^{\dagger}=\rho(h)\rho(g^{\prime})\rho^{\dagger}(h)=\rho(hg'h^{-1})\,.
\end{equation}
As regards the concerned simple flavor symmetry group $A_4$, it is enough and sufficient to impose the consistency condition of Eq.~\eqref{eq:consistency} on the group generators
\begin{equation}
\label{eq:cons_cond_A4}X\rho^*(S)X^{\dagger}=\rho(S^{\prime}),\qquad X\rho^*(T)X^{\dagger}=\rho(T^{\prime}),\qquad S',T'\in A_4\,,
\end{equation}
where the elements $S'$ and $T'$ should be of order two and three respectively. As a consequence, $S'$ belongs to the conjugacy class $3C_2$ and $T'$ belongs to $4C_3$ or $4C'_3$, i.e.
\begin{equation}
S'\in3C_2,\qquad T'\in 4C_3\cup4C'_3
\end{equation}
Because the different possible values of $S'$ and $T'$ are related to $(S', T')=(S, T)$ or $(S', T')=(S, T^2)$ by group conjugation, essentially only two kinds of CP transformations could be defined in the context of $A_4$ flavor symmetry\footnote{$A_4$ is isomorphic to
$\Delta(3\cdot2^2)$. The possible CP transformations which can be consistently defined within the $\Delta(3n^2)$ flavor group series have been comprehensively analyzed in Ref.~\cite{Ding:2015rwa}.}. The first one is fixed by the following consistency conditions
\begin{equation}
\label{eq:CP_1st_cons}X^{0}\rho^*(S)X^{0\dagger}=\rho(S),\qquad X^{0}\rho^*(T)X^{0\dagger}=\rho(T^2)\,,
\end{equation}
We can easily check that Eq.~\eqref{eq:CP_1st_cons} is fulfilled for all the irreducible representations of $A_4$ with
\begin{equation}
\label{eq:CP_1st}X^{0}=1\,.
\end{equation}
Taking into account the flavor symmetry, we find the most general CP transformation is of the same form as the flavor symmetry transformation in our basis. This kind of CP transformation has been discussed in Ref.~\cite{Ding:2013bpa}, and a dynamical model with both $A_4$ and CP symmetries was constructed. In this case, the second column of the PMNS matrix is predicted to be trimaximal, and the Dirac as well as Majorana CP phases are trivial if the $A_4$ and CP symmetry is broken down to $Z_2\times CP$ in the neutrino sector.

The second type of CP transformation is related to the value $(S', T')=(S, T)$. For the triplet representation $\mathbf{3}$, the corresponding consistency equations are
\begin{equation}
 X^0_{\mathbf{3}}\rho^*_{\mathbf{3}}(S)X^{0\dagger}_{\mathbf{3}}=\rho_{\mathbf{3}}(S), \qquad
  X^0_{\mathbf{3}}\rho^*_{\mathbf{3}}(T)X^{0\dagger}_{\mathbf{3}}=\rho_{\mathbf{3}}(T)\,.
\end{equation}
Disregarding an overall irrelevant phase, $X^0_{\mathbf{3}}$ is  determined to be
\begin{equation}
X^0_{\mathbf{3}}=\left(\begin{array}{ccc}
1 &~ 0 &~ 0 \\
0 &~ 0 &~ 1 \\
0 &~ 1 &~ 0 \\
\end{array}\right) \,,
\end{equation}
which is exactly the $\mu-\tau$ reflection symmetry~\cite{Harrison:2002kp,Grimus:2003yn,Farzan:2006vj}. A generic triplet field $\varphi_{\mathbf{3}}$ transforms under this CP symmetry as follows
\begin{equation}
\label{eq:CP_trans_3} \varphi_{\mathbf{3}}\rightarrow X^0_{\mathbf{3}}\varphi^{*}_{\mathbf{3}}=\left(\begin{array}{ccc}
1  &~  0  &~  0 \\
0  &~  0  &~  1 \\
0  &~  1  &~  0
\end{array}\right)\varphi^{*}_{\mathbf{3}}\,.
\end{equation}
However, it can be checked that the consistency conditions in Eq.~\eqref{eq:cons_cond_A4} are not satisfied separately for the nontrivial singlets $\mathbf{1}'$ and $\mathbf{1}''$ in the case of $(S', T')=(S, T)$. In order to resolve this issue, we consider a multiplet $\varphi\equiv(\varphi_{\mathbf{1}'}, \varphi_{\mathbf{1}''})$. The representation matrices of $A_4$ elements on the space of $\varphi$ are given by
\begin{equation}
\rho(S)=\begin{pmatrix}
1  &~  0  \\
0  &~  1
\end{pmatrix},\qquad \rho(T)=\begin{pmatrix}
\omega^2  &~  0  \\
0  &~  \omega
\end{pmatrix}\,.
\end{equation}
Then the solution for the consistency equation of Eq.~\eqref{eq:cons_cond_A4} exits, and it take the form
\begin{equation}
X_{\mathbf{1}^{\prime}, \mathbf{1}^{\prime\prime}}=\left(\begin{array}{cc}
0  &~  1  \\
1  &~   0
\end{array}\right)\,.
\end{equation}
Consequently the transformation rule of the singlet fields $\varphi_{\mathbf{1}^{\prime}}$ and $\varphi_{\mathbf{1}^{\prime\prime}}$ under this generalized CP is
\begin{equation}
\label{eq:CP_trans_singlet}\varphi_{\mathbf{1}'}\rightarrow\varphi^{*}_{\mathbf{1}''}, \quad  \varphi_{\mathbf{1}''}\rightarrow\varphi^{*}_{\mathbf{1}'}\,.
\end{equation}
Hence we conclude that the fields transforming as $\mathbf{1}'$ and $\mathbf{1}''$ should be present in pair or completely absent if one intends to implement this second kind of CP symmetry in a concrete model. Before ending this section, we would like to emphasize that the explicit form of the CP transformation depends on the chosen basis. In the frequently used Ma-Rajasekaran basis~\cite{Ma:2001dn}, $S$ and $T$ in the triplet $\mathbf{3}$ are represented by
\begin{equation}
S=\begin{pmatrix}
1  &~  0  &~  0 \\
0  &~  -1 &~  0  \\
0  &~  0  &~  -1
\end{pmatrix},\qquad T=\begin{pmatrix}
0  &~  1   &~   0  \\
0  &~  0  &~  1 \\
1  &~  0  &~  0
\end{pmatrix}\,.
\end{equation}
In the same manner, we find in this basis the first CP symmetry is
\begin{equation}
\label{eq:CP_first_Ma}
\varphi_{\mathbf{3}}\rightarrow\begin{pmatrix}
1 &~  0  &~  0\\
0 &~  0  &~  1 \\
0 &~  1  &~  0
\end{pmatrix}\varphi^{*}_{\mathbf{3}},\quad \varphi_{\mathbf{1}'}\rightarrow \varphi^{*}_{\mathbf{1}'},\quad \varphi_{\mathbf{1}''}\rightarrow \varphi^{*}_{\mathbf{1}''}\,,
\end{equation}
and the second is given by
\begin{equation}
\label{eq:CP_second_Ma}
\varphi_{\mathbf{3}}\rightarrow
\varphi^{*}_{\mathbf{3}},\quad \varphi_{\mathbf{1}'}\rightarrow \varphi^{*}_{\mathbf{1}''},\quad \varphi_{\mathbf{1}''}\rightarrow \varphi^{*}_{\mathbf{1}'}\,.
\end{equation}
These transformation rules are necessary and quite useful when one builds an $A_4$ model with CP symmetry in the Ma-Rajasekaran basis. In the following, we shall study the phenomenological predictions of this CP symmetry for lepton flavor mixing in a model independent way, both atmospheric mixing angle and the Dirac phases are predicted to be maximal. Furthermore, we shall construct a model to naturally realize these interesting model independent predictions.

\section{\label{sec:general_analysis}Lepton mixing from $A_4$ and CP symmetry breaking}

The first CP symmetry for $(S', T')=(S, T^2)$ given in Eq.~\eqref{eq:CP_1st} can be imposed on a generic $A_4$ model regardless of the matter content. The different mixing patterns that can be obtained from this CP symmetry and $A_4$ flavor group have been studied in Ref.~\cite{Ding:2013bpa}. In the present work, we shall be concerned with the second CP symmetry given by Eqs.~(\ref{eq:CP_trans_3}, \ref{eq:CP_trans_singlet}) for the case of $(S', T')=(S, T)$. The three generations of the left-handed leptons form an irreducible three-dimensional representation $\mathbf{3}$ of $A_4$. The $A_4$ and CP symmetry is broken to an abelian subgroup $G_{l}$ and to $G_{\nu}$ in the charged lepton and neutrino sectors, respectively. $G_{l}$ is required to be able to distinguish the three generations of charged leptons, consequently this group should have at least three different elements with non-degenerate eigenvalues. Therefore $G_{l}$ can be either
a $Z_3$ or a Klein group. In the neutrino sector, we shall consider
the residual symmetry $G_{\nu}$ is $Z_2\times CP$. The group $A_4$ has three $Z_2$ subgroups $Z^{S}_2=\left\{1, S\right\}$, $Z^{TST^2}_{2}=\left\{1, TST^2\right\}$, $Z^{T^2ST}_{2}=\left\{1, T^2ST\right\}$, four $Z_3$ subgroups $Z^{T}_3=\left\{1,T, T^2\right\}$, $Z^{ST}_3=\left\{1, ST, T^2S\right\}$, $Z^{TS}_3=\left\{1, TS, ST^2\right\}$, $Z^{STS}_3=\left\{1, STS, ST^2S\right\}$ and a unique Klein group $K_{4}=\left\{1, S, TST^2, T^2ST\right\}$. In the case of $G_{l}=K_4$, one column of the PMNS matrix would be $(1,0,0)^{T}$ which is not compatible with the data at $3\sigma$ level. As a consequence, we shall study the case in which $G_{l}$ is a $Z_3$ subgroup of $A_4$. Furthermore, it is sufficient to consider only the residual symmetries $G_{l}=Z^{T}_{3}$ and $G_{\nu}=Z^{S}_2\times CP$, since other possible choices of $G_{l}$ and $G_{\nu}$ are related by similarity transformations to this representative one and thus don't
lead to new mixing patterns.

The residual group $Z^{T}_3$ leads to the following constraint on the charged lepton mass matrix $m_{l}$,
\begin{equation}\label{eq:Z3T_cons}
\rho^{\dagger}_{\mathbf{3}}(T)m^{\dagger}_{l}m_{l}\rho_{\mathbf{3}}(T)=m^{\dagger}_{l}m_{l}\,,
\end{equation}
where $m_l$ is written in the convention with right-handed charged leptons on the left-hand side and left-handed leptons on the right-hand side. Since the representation matrix $\rho_{\mathbf{3}}(T)=\text{diag}\left(1, \omega^2, \omega\right)$ is diagonal in our working basis, the hermitian combination $m^{\dagger}_{l}m_{l}$ is also diagonal, i.e.
\begin{equation}
m^{\dagger}_{l}m_{l}=\text{diag}\left(m^2_{e},m^2_{\mu},m^2_{\tau}\right)\,,
\end{equation}
where $m_e$, $m_{\mu}$ and $m_{\tau}$ are the electron, muon and tau masses, respectively.

The residual symmetry of the neutrino sector is the direct product of the $Z^{S}_2$ subgroup and a CP symmetry which is represented by a three-by-three matrix $X_{\nu}$. This residual symmetry is consistently defined if and only if the condition
\begin{eqnarray}\label{eq:consistent-two}
X_{\nu}\rho_{\mathbf{3}}^{*}(S)X_{\nu}^{-1}=\rho_{\mathbf{3}}(S)
\end{eqnarray}
is fulfilled. We find that four out of the twelve generalized CP transformations are acceptable,
\begin{equation}
\label{eq:X_nu}
X_{\nu}=X^{0}_{\mathbf{3}},~\rho_{\mathbf{3}}(S)X^{0}_{\mathbf{3}},~\rho_{\mathbf{3}}(TST^2)X^{0}_{\mathbf{3}},~\rho_{\mathbf{3}}(T^2ST)X^{0}_{\mathbf{3}}\,.
\end{equation}
Preserving the residual symmetry $Z^{S}_2\times CP$ in the neutrino sector requires that the neutrino mass matrix $m_{\nu}$ should satisfy the conditions
\begin{equation}
\label{eq:mnu_inv_cond}\rho_{\mathbf{3}}^{T}(S)m_{\nu}\rho_{\mathbf{3}}(S)=m_{\nu},\qquad X_{\nu}^Tm_{\nu}X_{\nu}=m^{*}_{\nu}\,.
\end{equation}
The most general neutrino mass matrix invariant under the residual flavor symmetry $Z^{S}_2$ takes the form
\begin{eqnarray}\label{eq:general_mass_matrix}
m_{\nu}=\alpha \left(\begin{array}{ccc}
   2  &  -1  &  -1  \\
   -1  &  2  &  -1  \\
   -1  &  -1  &  2
 \end{array}\right)+\beta \left(\begin{array}{ccc}
    1  &  0  &  0  \\
    0  &  0  &  1  \\
    0  &  1  &  0
  \end{array}\right)+\gamma \left(\begin{array}{ccc}
    0  &  1  &  1  \\
    1  &  1  &  0  \\
    1  &  0  &  1
  \end{array}\right)+\delta \left(\begin{array}{ccc}
    0  &  1  &  -1  \\
    1  &  -1  &  0  \\
    -1  &  0  &  1
  \end{array}\right)\,,
\end{eqnarray}
where $\alpha$, $\beta$, $\gamma$ and $\delta$ are generally complex parameters, and they are further constrained to be real or
pure imaginary by the residual CP symmetry $X_{\nu}$. Subsequently a tri-bimaximal transformation is performed on the light neutrino fields, then $m_{\nu}$ becomes
\begin{eqnarray}
\label{eq:general_mass_matrix_two}
m_{\nu}^{\prime}=U_{TB}^{T}m_{\nu}U_{TB}= \left(
\begin{array}{ccc}
 3 \alpha +\beta -\gamma  &~ 0 ~& -\sqrt{3} \delta  \\
 0 &~ \beta +2 \gamma  ~& 0 \\
 -\sqrt{3} \delta  &~ 0 ~& 3 \alpha -\beta +\gamma  \\
\end{array}
\right)\,,
\end{eqnarray}
where $U_{TB}$ is the prominent tri-bimaximal mixing matrix,
\begin{equation}
U_{TB}=\left(\begin{array}{ccc}
\sqrt{\frac{2}{3}}  &~   \frac{1}{\sqrt{3}}  ~&  0  \\
-\frac{1}{\sqrt{6}}  &~  \frac{1}{\sqrt{3}}  ~&  -\frac{1}{\sqrt{2}}  \\
-\frac{1}{\sqrt{6}}  &~  \frac{1}{\sqrt{3}}  ~&  \frac{1}{\sqrt{2}}
\end{array}\right)\,.
\end{equation}
Since $m'_{\nu}$ is a block diagonal symmetric matrix , it can be easily diagonalized as
\begin{equation}
U'^{T}_{\nu}m^{\prime}_{\nu}U^{\prime}_{\nu}=\text{diag}(m_{1},m_{2},m_{3})\,,
\end{equation}
where $m_{1,2,3}$ are the light neutrino masses. The explicit form of the unitary matrix $U'_{\nu}$ would be presented for different choices of $X_{\nu}$ in the following. Because $X_{\nu}$ and $\rho_{\mathbf{3}}(S)X_{\nu}$ lead to the same constraint on the neutrino mass matrix, the four admissible residual CP symmetries in Eq.~\eqref{eq:X_nu} fall into two categories.

\begin{description}[labelindent=-0.6em, leftmargin=0.2em]
\item[~~(\uppercase\expandafter{\romannumeral1})] $X_{\nu}=X^{0}_{\mathbf{3}},~\rho_{\mathbf{3}}(S)X^{0}_{\mathbf{3}}$

In this case, the parameters $\alpha$, $\beta$ and $\gamma$ are restricted to be real and $\delta$ is pure imaginary. The unitary matrix $U^{\prime}_{\nu}$ is determined to be
\begin{equation}
U^{\prime}_{\nu}=\left(\begin{array}{ccc}
 \cos\theta  &~    0   ~&    \sin\theta   \\
0   &~  1   ~&   0   \\
-i\sin\theta   &~  0  ~&   ~i\cos\theta
\end{array}\right)K_{\nu}\,,
\end{equation}
where $K_{\nu}$ is a diagonal matrix with entries equal to $\pm1$ and $\pm i$, and it is necessary to make the neutrino masses $m_{1,2,3}$ positive. The rotation angle $\theta$ is given by
\begin{equation}
\label{eq:theta_caseI}
\tan2\theta=\frac{i\delta}{\sqrt{3}\,\alpha}\,.
\end{equation}
The three light neutrino masses are
\begin{equation}
m_{1}=\left|\beta-\gamma+\frac{3\alpha}{\cos2\theta}\right|\,, \quad m_{2}=\frac{1}{2}\left|\beta+2\gamma\right|\,, \quad m_{3}=\left|\beta-\gamma-\frac{3\alpha}{\cos2\theta}\right|\,.
\end{equation}
One sees that the neutrino masses can be either normal ordering (NO) or inverted ordering (IO). As the charged lepton mass matrix $m^{\dagger}_{l}m_{l}$ is diagonal, there is no contribution to the lepton flavor mixing from the diagonalization of charged lepton mass matrix, and the PMNS matrix takes the form
\begin{equation}
\label{eq:PMNS_I}
U_{PMNS}=U_{TB}U^{\prime}_{\nu}=\frac{1}{\sqrt{6}}
\begin{pmatrix}
 2 \cos \theta  &~ \sqrt{2} ~& 2 \sin \theta  \\
 -\cos \theta+i\sqrt{3}\sin\theta   &~ \sqrt{2} ~& -\sin\theta-i\sqrt{3}\cos\theta    \\
 -\cos \theta-i\sqrt{3}\sin\theta   &~ \sqrt{2} ~& -\sin\theta+i\sqrt{3}\cos\theta
\end{pmatrix}K_{\nu}\,.
\end{equation}
This mixing pattern can also be obtained from $S_4$ group combined with CP symmetry~\cite{Feruglio:2012cw,Ding:2013hpa}. The three lepton mixing angles can be read off as
\begin{equation}\label{eq:mixing_angles}
\sin^2\theta_{13}=\frac{2}{3}\sin^2\theta,\quad \sin^2\theta_{12}=\frac{1}{2+\cos2\theta}, \quad \sin^2\theta_{23}=\frac{1}{2} \,.
\end{equation}
Obvious the atmospheric mixing angle $\theta_{23}$ is maximal, and the solar and the reactor mixing angles satisfy the following sum rule
\begin{equation}
\label{eq:correlation_trimax}3\sin^2\theta_{12}\cos^2\theta_{13}=1\,.
\end{equation}
The best fit value $\sin^2\theta_{13}=0.0234$~\cite{Capozzi:2013csa}
can be accommodated for $\theta\simeq0.06\pi$. Accordingly the solar mixing angle is $\sin^2\theta_{12}\simeq0.341$ which is in the experimentally preferred $3\sigma$ region~\cite{Capozzi:2013csa}. Moreover, both Majorana CP violating phases $\alpha_{21}$ and $\alpha_{31}$ are trivial, they are $0$ or $\pi$. The Jarlskog invariant $J_{CP}$ describing the CP violation takes a simple form
\begin{equation}
J_{CP}=-\frac{\sin2\theta}{6\sqrt{3}}\,.
\end{equation}
Consequently the Dirac CP phase $\delta_{CP}$ is predicted to be maximal with
\begin{equation}
\sin\delta_{CP}=\left\{\begin{array}{cc}
-1, ~&~ 0<\theta<\pi/2\,, \\  1, ~&~ \pi/2<\theta<\pi\,.
\end{array}\right.
\end{equation}
In light of the weak evidence of $\delta_{CP}\simeq3\pi/2$ from T2K~\cite{Abe:2015awa}, this mixing pattern is quite interesting. Here the maximal atmospheric mixing and maximal Dirac phase are due to the presence of $\mu-\tau$ reflection symmetry $X^{0}_{\mathbf{3}}$~\footnote{It was shown that maximal $\theta_{23}$ and $\delta_{CP}$ could follow
from some general assumptions without imposing a CP symmetry while the Majorana phases are not constrained~\cite{Joshipura:2015dsa}.}, and correlation in Eq.~\eqref{eq:correlation_trimax} arises from the remnant flavor symmetry $Z^{S}_2$. Notice that the $\mu-\tau$ reflection symmetry restricts neither the reactor mixing angle nor the solar mixing angle. The neutrinoless double beta (0$\nu\beta\beta$) decay processes $(A,Z)\rightarrow(A, Z+2)+2e^{-}$ is important to test the Majorana nature of neutrinos. Many new experiments are currently running, under construction, or in the planing phase. The sensitivity to this rare process would be increased significantly in future. Besides the nuclear matrix elements, the amplitude of the $0\nu\beta\beta$ decay is proportional to the quantity~\cite{Agashe:2014kda}
\begin{equation}
m_{ee}=\left|\sum_{i}m_iU^{2}_{PMNS,1i}\right|\,,
\end{equation}
which is known as the effective Majorana neutrino mass for $0\nu\beta\beta$ decay. The predictions for $m_{ee}$ strongly depend on the type of the neutrino mass spectrum. One can express $m_{ee}$ as a function of the lightest neutrino mass $m_{min}$, the oscillation mass splittings $\delta m^2\equiv m^2_{2}-m^2_{1}$ and $\Delta m^2\equiv m^2_3-(m^2_1+m^2_{2})/2$~\cite{Capozzi:2013csa} and the neutrino mixing matrix elements. For NO neutrino mass spectrum, one has
\begin{equation}
m_{1}=m_{min}, \quad m_{2}=\sqrt{m^2_{min}+\delta m^2}, \quad m_{3}=\sqrt{m^2_{min}+\delta m^2/2+\Delta m^2}\,,
\end{equation}
while in the case of IO,
\begin{equation}
m_{1}=\sqrt{m^2_{min}-\delta m^{2}/2-\Delta m^{2}}, \quad m_{2}=\sqrt{m^2_{min}+\delta m^{2}/2-\Delta m^{2}}, \quad   m_{3}=m_{min}\,.
\end{equation}
For the predicted mixing pattern in Eq.~\eqref{eq:PMNS_I}, the effective mass $m_{ee}$ is of the form
\begin{equation}
m_{ee}=\frac{1}{3}\left|2m_{1}\cos^{2}\theta+k_{1}m_{2}+2k_{2}m_{3}\sin^{2}\theta\right|\,,
\end{equation}
where $k_{1},k_{2}=\pm1$ originate from the CP parity matrix $K_{\nu}$. The possible values of $m_{ee}$ with respect the lightest neutrino mass $m_{min}$ are displayed in figure~\ref{fig:mee}, where the neutrino mass squared differences $\delta m^2$ and $\Delta m^2$ freely vary within their $3\sigma$ intervals and the parameter $\theta$ varies in the interval $0\leq\theta\leq\pi$ while the three mixing angles are required to be in the experimentally preferred $3\sigma$ ranges~\cite{Capozzi:2013csa}. We see that $m_{ee}$ is around 0.016eV and 0.050eV for $(k_1, k_2)=(+, +), (+, -)$ and $(k_1, k_2)=(-, +), (-, -)$ respectively in the case of IO. The next generation $0\nu\beta\beta$ decay experiments will be able to probe the full IO region and these predictions can be tested. The effective mass $m_{ee}$ depends on $m_{min}$ and the CP parities in case of NO, it has a lower limit $m_{ee}\geq0.004$ eV and $m_{ee}\geq0.0012$ eV for $(k_1, k_2)=(+, +)$ and $(+, -)$ respectively, and $m_{ee}$ can be strongly suppressed to be smaller than $10^{-4}$ eV for $(k_1, k_2)=(-, +), (-, -)$.

\begin{figure}[t!]
\centering
\includegraphics[width=0.65\textwidth]{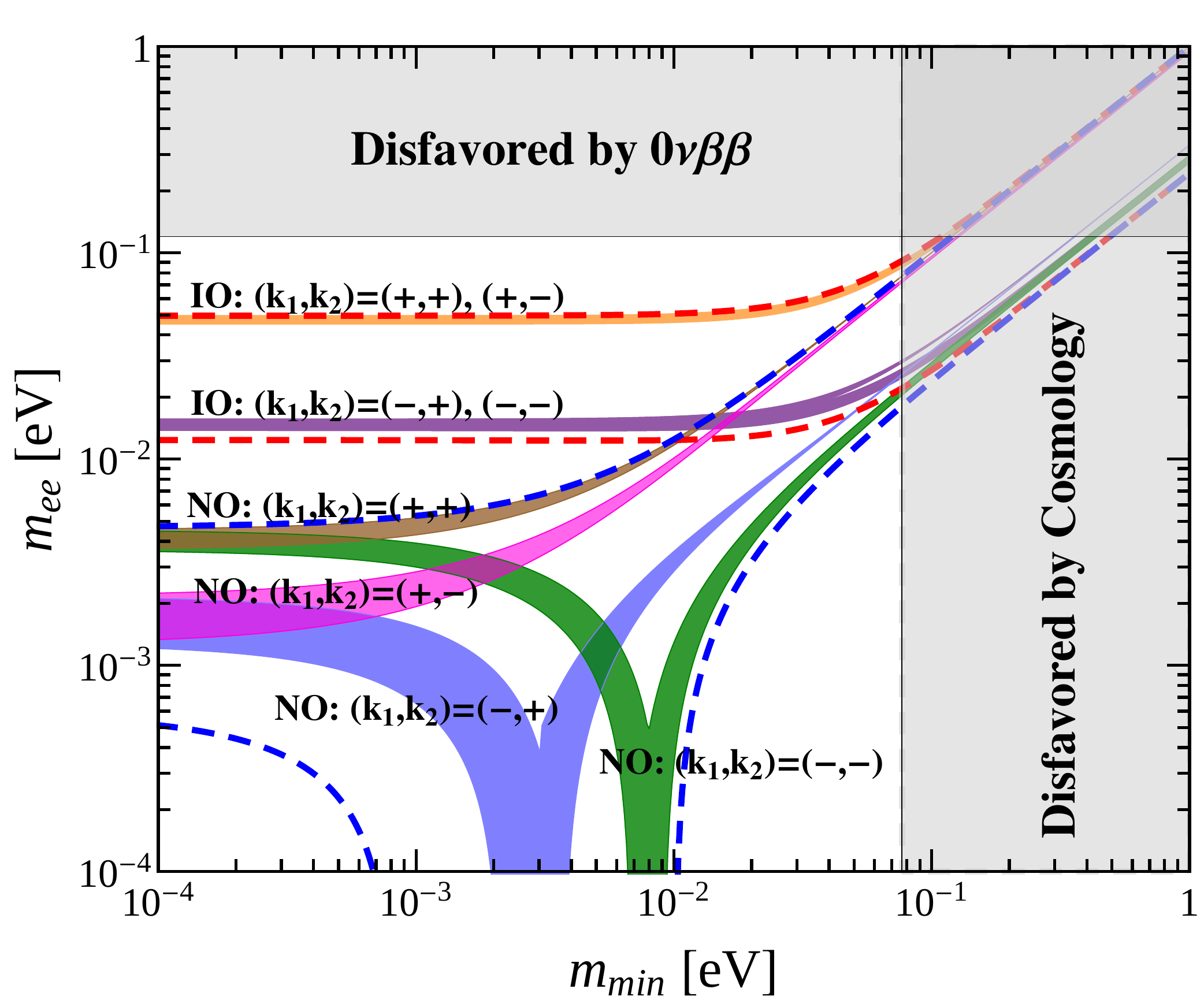}
\caption{\label{fig:mee} The effective Majorana mass $m_{ee}$ as a function of the lightest neutrino mass $m_{min}$ for different CP parities. The red (blue) dashed lines indicate the most general allowed regions for IO (NO) spectrum obtained by varying all the neutrino oscillation parameters over their $3\sigma$ ranges~\cite{Capozzi:2013csa}. The horizontal grey band denotes the present most stringent upper bound $|m_{ee}|<0.120$ eV from EXO-200~\cite{Auger:2012ar, Albert:2014awa} and KamLAND-ZEN~\cite{Gando:2012zm}. The vertical grey exclusion band is the current limit on $m_{min}$ from the cosmological data of $\sum m_i<0.230$ eV by the Planck collaboration~\cite{Ade:2013zuv} at $95\%$ confidence level.}
\end{figure}

\item[~~(\uppercase\expandafter{\romannumeral2})] $X_{\nu}=\rho_{\mathbf{3}}(TST^2)X^{0}_{\mathbf{3}}$, $\rho_{\mathbf{3}}(T^2ST)X^{0}_{\mathbf{3}}$

Invariance of the neutrino mass matrix $m_{\nu}$ under these residual CP transformations implies that $\beta$ and $\gamma$ are real while $\alpha$ and $\delta$ are pure imaginary. The neutrino mass matrix $m'_{\nu}$ is diagonalized by the following unitary transformation
\begin{equation}
U^{\prime}_{\nu}=\frac{1}{\sqrt{2}}\left(
\begin{array}{ccc}
 e^{-\frac{i \theta}{2}} &~ 0 ~& e^{-\frac{i \theta}{2}} \\
 0 &~ \sqrt{2} ~& 0 \\
 -i e^{\frac{i \theta }{2}} &~ 0 ~& i e^{\frac{i \theta }{2}} \\
\end{array}
\right)K_{\nu}\,,
\end{equation}
with
\begin{equation}
\tan\theta=\frac{3i\alpha}{\gamma-\beta}\,.
\end{equation}
Consequently the PMNS mixing matrix takes the form,
\begin{equation}\label{eq:PMNS_II}
U_{PMNS}=\frac{1}{2\sqrt{3}}\left(
\begin{array}{ccc}
e^{-\frac{i \theta}{2}} \left(-1+i \sqrt{3} e^{i \theta }\right) ~&~ 2 ~&~ e^{-\frac{i \theta}{2} } \left(-1-i \sqrt{3} e^{i \theta }\right) \\
 e^{-\frac{i \theta}{2}} \left(-1-i \sqrt{3} e^{i \theta }\right) ~&~ 2 ~&~ e^{-\frac{i \theta}{2} } \left(-1+i \sqrt{3} e^{i \theta }\right) \\
  2 e^{-\frac{i \theta}{2}} ~&~ 2 ~&~ 2 e^{-\frac{i \theta}{2} } \\
\end{array}
\right)\,.
\end{equation}
We can extract the mixing angles from Eq.~\eqref{eq:PMNS_II} and find
\begin{equation}
\sin^2\theta_{12}=\frac{2}{4+\sqrt{3} \sin \theta }, \quad \sin^2\theta_{23}=\frac{2+\sqrt{3} \sin \theta }{4+\sqrt{3} \sin \theta }\,. 
\end{equation}
For the CP invariants we have
\begin{eqnarray}
\label{eq:mixing2}&& |J_{CP}|=\frac{1}{6 \sqrt{3}}|\cos \theta |, \quad |I_{1}|=\frac{1}{18}|\sqrt{3}+2 \sin \theta |, \quad |I_{2}|=\frac{1}{6 \sqrt{3}}|\cos \theta |\,,
\end{eqnarray}
where the invariants $I_1$ and $I_2$ are defined for the Majorana phases
\begin{equation}
\begin{split}
I_1&=\Im\left(U^{*2}_{PMNS,11}U^{2}_{PMNS,12}\right)=\cos^2\theta_{12}\sin^2\theta_{12}\cos^4\theta_{13}\sin\alpha_{21},\\
I_2&=\Im\left(U^{*2}_{PMNS,11}U^{2}_{PMNS,13}\right)=\cos^2\theta_{12}\cos^2\theta_{13}\sin^2\theta_{13}\sin\left(\alpha_{31}-2\delta_{CP}\right)\,.
\end{split}
\end{equation}
Expressing the parameter $\theta$ in terms of $\theta_{13}$, we find the following sum rules among the lepton mixing angles,
\begin{equation}
\sin^2\theta_{12}\cos^2\theta_{13}=\frac{1}{3}, \qquad  \sin^2\theta_{23}=\frac{1}{3}(2-\tan^2\theta_{13})\,.
\end{equation}
The reactor mixing angle $\sin^2\theta_{13}$ is minimized for $\theta=\pi/2$, and accordingly we obtain
\begin{equation}
\begin{split}
\sin^2\theta_{13}\Big|_{\theta=\pi/2}&=\frac{2-\sqrt{3}}{6}\simeq0.0447\,, \\
\sin^2\theta_{12}\Big|_{\theta=\pi/2}&=\frac{2}{4+\sqrt{3}}\simeq0.349\,,\\ \sin^2\theta_{23}\Big|_{\theta=\pi/2}&=\frac{2+\sqrt{3}}{4+\sqrt{3}}\simeq0.651\,,\\
\sin\delta_{CP}\Big|_{\theta=\pi/2}&=\cos\alpha_{21}\Big|_{\theta=\pi/2}=\sin\alpha_{31}\Big|_{\theta=\pi/2}=0\,.
\end{split}
\end{equation}
As both $\theta_{13}$ and $\theta_{23}$ are outside the present $3\sigma$ ranges~\cite{Capozzi:2013csa}, this mixing pattern isn't phenomenologically viable unless higher order corrections could lead to the agreement with experimental data in a model.

\end{description}

\section{\label{sec:model}The structure of the model}

\begin{table} [t!]
\begin{center}
\renewcommand{\tabcolsep}{2.0mm}
{\footnotesize
\begin{tabular}{|c||c|c|c|c|c|c||c|c|c|c|c|c|c||c|c|c|c|c|c|c|}
\hline\hline

 {\tt Field}& $l$ &  $\nu^c$ &  $e^c$ & $\mu^c$ & $\tau^c$ &  $h_{u,d}$ & $\zeta^{\prime}$ & $\zeta^{\prime\prime}$ &
$\varphi_T$ & $\xi$ &  $\varphi_S$ & $\sigma^{\prime}$ & $\sigma^{\prime\prime}$ & $\zeta^0$ & $\varphi^0_T$   & $\xi^0$ & $\sigma^{0}$ & $\varphi^0_S$     \\ \hline
$A_4$ & $\mathbf{3}$ & $\mathbf{3}$ & $\mathbf{1}$ & $\mathbf{1}$ &
$\mathbf{1}$ &   $\mathbf{1}$   & $\mathbf{1}^{\prime}$ & $\mathbf{1}^{\prime\prime}$ &  $\mathbf{3}$
& $\mathbf{1}$ & $\mathbf{3}$ & $\mathbf{1}^{\prime}$ & $\mathbf{1}^{\prime\prime}$ & $\mathbf{1}$   &   $\mathbf{3}$  & $\mathbf{1}$ &   $\mathbf{1}$ & $\mathbf{3}$  \\
\hline

$Z_4$ & $-1$ & $-1$ & $-i$ & 1 & $i$ &   1 & $i$ & $i$ & $i$ & $1$ & $1$ & $1$ & $1$ & $-1$ & $-1$  &  $1$ & $1$ & $1$    \\ \hline

$Z_5$  &  $\omega_5$  &  $\omega^4_5$  &  $\omega^4_5$  &  $\omega^4_5$  & $\omega^4_5$  &    1  & 1 &  1 & $1$ & $\omega^2_5$  &  $\omega^2_5$ &  $\omega_5$ &  $\omega_5$ &
  $1$ & $1$ &  $\omega^3_5$  & $\omega_5$  & $\omega^2_5$    \\ \hline

$U(1)_R$ & $1$& $1$ & $1$ & $1$ & $1$ &   $0$ & $0$ & $0$& $0$  & $0$ & $0$& $0$ & $0$ & $2$ & $2$ &  $2$ & $2$ & $2$  \\ \hline \hline
\end{tabular}}
\caption{\label{tab:rule} Transformation properties of the matter fields, flavon fields and driving fields under the flavor symmetry $A_4 \times Z_4\times Z_5$ and $U(1)_R$, where $\omega_5$ is the fifth root of unit $\omega_5=e^{2\pi i/5}$. }
\end{center}
\end{table}

In this section, we shall construct a model to realize the interesting mixing pattern of case I. We will formulate our model in the context of the minimal supersymmetric standard model and all supersymmetry breaking effects are neglected in the following. The $A_4$ flavor symmetry as well as the CP symmetry defined in Eqs.~(\ref{eq:CP_trans_3}, \ref{eq:CP_trans_singlet}) are imposed at higher energy scale. The auxiliary symmetry is chosen to be $Z_4\times Z_5$ in order to eliminate unwanted operators, to ensure the
needed vacuum alignment and to reproduce the observed charged lepton mass hierarchies. We assign the three generations of left-handed lepton doublets $l$ and right-handed neutrino $\nu^{c}$ to $A_{4}$ triplet $\mathbf{3}$, while the right-handed charged leptons $e^{c}$, $\mu^{c}$ and $\tau^{c}$ transform as singlet $\mathbf{1}$ under $A_4$. A $U(1)_R$ symmetry related to the usual $R-$parity and the presence of driving fields are common features of supersymmetric flavor models. The flavon fields, matter fields and driving fields carry zero, one and two unit of $R$ charges respectively. We summarize the field content of the model and the symmetry assignments in table~\ref{tab:rule}. Notice that the flavons $\zeta'$ and $\zeta''$ have the same quantum numbers of $Z_4\times Z_5$ and $U(1)_{R}$, and they transform as $\zeta'\rightarrow\zeta''^{*}$ and $\zeta''\rightarrow\zeta'^{*}$ under the action of the CP symmetry. The similar hold true for the flavon fields $\sigma^{\prime}$ and $\sigma^{\prime\prime}$. In our model, the lepton mixing matrix is exactly the famous tri-bimaximal mixing at leading order (LO), and a non-zero reactor mixing angle $\theta_{13}$ originates from the next-to-leading order (NLO) corrections. As a result, $\theta_{13}$ is naturally smaller than the other two mixing angles $\theta_{12}$ and $\theta_{23}$. The correct size of $\theta_{13}$ can be achieved in the model since the NLO contributions are suppressed by a factor of order $0.1\sim0.2$ with respect to the LO ones.

\subsection{\label{sec:vacuum_alignment_model2}Vacuum alignment}

All the flavon fields of our model can be divided into two sets $\{\zeta', \zeta^{''}, \varphi_T \}$ and $\{\xi, \varphi_{S},\sigma^{\prime}, \sigma^{\prime\prime}\}$ which enter into the charged lepton and neutrino mass terms respectively at LO. The driving superpotential for $\zeta'$, $\zeta^{''}$ and $\varphi_T$ is given by
\begin{equation}
\label{eq:driving_pote_clepton}
w^{l}_d=f_1\zeta^{0}\zeta^{\prime}\zeta^{\prime\prime}+f_{2}\zeta^{0}\left(\varphi_T\varphi_T\right)_{\mathbf{1}}
+f_{3}\zeta^{\prime}\left(\varphi^0_T\varphi_T\right)_{\mathbf{1}^{\prime\prime}}+f_{4}\zeta^{\prime\prime}\left(\varphi^0_T\varphi_T\right)_{\mathbf{1}^{\prime}}
+f_5\left(\varphi^0_T(\varphi_T\varphi_T)_{\mathbf{3}_{S}}\right)_{\mathbf{1}}\,,
\end{equation}
where $(\ldots)_{\bf{r}}$ denotes a contraction into the irreducible representation $\mathbf{r}$. Note that we have neglected the term $\left(\varphi^0_T(\varphi_T\varphi_T)_{\mathbf{3}_{A}}\right)_{\mathbf{1}}$ which vanishes due to the antisymmetric property of the contraction $(\varphi_T\varphi_T)_{\mathbf{3}_{A}}$. As we assume the theory is invariant under the CP symmetry in Eqs.~(\ref{eq:CP_trans_3}, \ref{eq:CP_trans_singlet}), the coupling constants $f_{1}$, $f_{2}$ and $f_{5}$ should be real while $f_3$ and $f_4$ are generally complex numbers with $f_{3}=f^{*}_{4}$. The driving fields are assumed to have vanishing vacuum expectation values (VEVs). In the exact supersymmetric limit, the $F-$terms of the driving fields have to vanish at the minimum of the scalar potential such that the vacuum of the flavon fields is aligned. Then the $F-$term conditions obtained from the driving fields $\zeta^0$ and $\varphi^0_T$ read as
\begin{equation}
\begin{split}
\frac{\partial
w^{l}_d}{\partial\zeta^0}&=f_1\zeta^{\prime}\zeta^{\prime\prime}+f_2\left(\varphi^2_{T_1}+2\varphi_{T_2}\varphi_{T_3}\right)=0\,,\\
\frac{\partial
w^{l}_d}{\partial\varphi^0_{T_1}}&=f_3\zeta^{\prime}\varphi_{T_3}+f_4\zeta^{\prime\prime}\varphi_{T_2}+\frac{2}{3}f_5\left(\varphi^2_{T_1}-\varphi_{T_2}\varphi_{T_3}\right)=0\,,\\
\frac{\partial
w^{l}_d}{\partial\varphi^0_{T_2}}&=f_3\zeta^{\prime}\varphi_{T_2}+f_4\zeta^{\prime\prime}\varphi_{T_1}+\frac{2}{3}f_5\left(\varphi^2_{T_2}-\varphi_{T_1}\varphi_{T_3}\right)=0\,,\\
\frac{\partial
w^{l}_d}{\partial\varphi^0_{T_3}}&=f_3\zeta^{\prime}\varphi_{T_1}+f_4\zeta^{\prime\prime}\varphi_{T_3}+\frac{2}{3}f_5\left(\varphi^2_{T_3}-\varphi_{T_1}\varphi_{T_2}\right)=0\,.
\end{split}
\end{equation}
These equations admit a nontrivial solution
\begin{equation}\label{eq:vev_charged}
\langle\zeta^{\prime}\rangle=v_{\zeta},\quad     \langle\zeta^{\prime\prime}\rangle=0 , \qquad \langle\varphi_T\rangle=\left(\begin{array}{c}
0\\v_T\\0  \end{array}\right)\,,
\end{equation}
with the condition
\begin{equation}
v_T=-\frac{3f_3}{2f_5}v_{\zeta}\,,
\end{equation}
where $v_{\zeta}$ is a undetermined complex parameter. The coupling constants $f_3$ and $f_5$ naturally have absolute values of order one, consequently the VEVs $v_{\zeta}$ and $v_{T}$ are of the same order of magnitude. In order to generate the observed mass hierarchy among the charged leptons, we choose
\begin{equation}\label{eq:hierarchy_ch}
|v_{\zeta}|,~ |v_T|\sim\lambda^2\Lambda\,,
\end{equation}
where $\lambda\simeq0.23$ is the size of the Cabibbo angle. For the flavon fields $\xi$, $\varphi_S$, $\sigma^{\prime}$ and $\sigma^{\prime\prime}$ in the neutrino sector, the LO driving superpotential $w^{\nu}_d$ takes the form
\begin{equation}\label{eq:driving_potential_nu}
w^{\nu}_d= M\xi^{0}\xi+g_{1}\xi^{0}\sigma^{\prime}\sigma^{\prime\prime}+g_{2}\sigma^{0}\xi^{2}+g_{3}\sigma^{0}\left(\varphi_S\varphi_S\right)_{\mathbf{1}}
+g_{4}\sigma^{\prime}\left(\varphi^{0}_S\varphi_S\right)_{\mathbf{1}^{\prime\prime}}+g_{5}\sigma^{\prime\prime}\left(\varphi^{0}_S\varphi_S\right)_{\mathbf{1}^{\prime}}\,,
\end{equation}
where the coupling $g_{4}$ and $g_{5}$ are general complex numbers with $g_{4}=g^{*}_{5}$ and the other couplings $g_i$ $(i=1,2,3)$ and the mass parameter $M$ are real because of the imposed CP symmetry. The equations for the vanishing of the derivatives of $w^{\nu}_d$ with respect to each component of the driving fields read as
\begin{eqnarray}
\nonumber&&\frac{\partial w^{\nu}_d}{\partial\xi^0}=M \xi+g_{1}\sigma^{\prime}\sigma^{\prime\prime}=0\,,\\
\nonumber&&\frac{\partial w^{\nu}_d}{\partial\sigma^0}=g_{2}\xi^2+g_{3}(\varphi_{S_1}^2+2\varphi_{S_2}\varphi_{S_3})=0\,,\\
\nonumber&&\frac{\partial w^{\nu}_d}{\partial\varphi^0_{S_1}}=g_{4}\sigma^{\prime}  \varphi_{S_3}+g_5 \sigma^{\prime\prime} \varphi_{S_2}=0\,, \\
\nonumber&&\frac{\partial w^{\nu}_d}{\partial\varphi^0_{S_2}}=
g_{4}\sigma^{\prime}  \varphi_{S_2}+g_5 \sigma^{\prime\prime} \varphi_{S_1}=0\,, \\
&&\frac{\partial w^{\nu}_d}{\partial\varphi^0_{S_3}}=g_{4}\sigma^{\prime}  \varphi_{S_1}+g_5 \sigma^{\prime\prime} \varphi_{S_3}=0\,.
\end{eqnarray}
These equations are satisfied by the alignment
\begin{equation}\label{eq:neutrino_vacuum}
\langle\xi\rangle=v_{\xi}, \qquad
\langle\varphi_S\rangle =\left(\begin{array}{c}
   1 \\
   1  \\
   1
  \end{array}\right)v_S, \qquad \langle\sigma^{\prime}\rangle=v_{\sigma}, \qquad
\langle\sigma^{\prime\prime}\rangle=-\frac{g_{4}}{g_{5}}v_{\sigma}\,,
\end{equation}
where the VEVs $v_{\xi}$, $v_S$ and $v_{\sigma}$ are related by
\begin{equation}\label{eq:VEVs_relation}
v^2_S=-\frac{g_2}{3g_{3}}v^2_{\xi},\qquad v^2_{\sigma}=\frac{g_{5}M}{g_{1}g_{4}}v_{\xi}\,,
\end{equation}
with $v_{\xi}$ being a free parameter which is in general complex. One sees that the phase difference between $v_{\xi}$ and $v_{S}$ is $0$, $\pi$ for $g_2g_3<0$ and $\pm\pi/2$ for $g_2g_3>0$ and the phase difference between $v_{\xi}$ and $v^2_{\sigma}$ is twice as large as the phase of $g_{5}$ up to $\pi$. As we shall show in section~\ref{subsec:model_LO}, the common phase of $v_{\xi}$ and $v_{S}$ can be factored out in the neutrino mass $m_{\nu}$ and consequently it can be absorbed by the redefining the lepton fields. Without loss of generality, we can take $v_{\xi}$ to be real. Then $v_{S}$ would be real for $g_{2}g_{3}<0$ and pure imaginary for $g_{2}g_{3}>0$ and $v^2_{\sigma}$ is generally a complex parameter. It is easy to check that the vacuum in Eq.~\eqref{eq:neutrino_vacuum} is stable under small perturbations. If one introduces small quantities in the VEVs of the flavons as follows
\begin{equation}\label{eq:neutrino_vacuum_perturbation}
\langle\varphi_S\rangle =\left(\begin{array}{c}
   1+\epsilon^{S}_1 \\
   1+\epsilon^{S}_2  \\
   1+\epsilon^{S}_3
  \end{array}\right)v_S, \qquad
  \langle\sigma^{\prime}\rangle=(1+\epsilon_{\sigma^{\prime}})\,v_{\sigma}, \qquad
\langle\sigma^{\prime\prime}\rangle=-(1+\epsilon_{\sigma^{\prime\prime}})g_{4}\,v_{\sigma}/g_{5} \,.
\end{equation}
After some straightforward algebraic calculations, one can show that
the only solution minimizing the scalar potential in the supersymmetric limit is given by $(\epsilon^{S}_1, \epsilon^{S}_2, \epsilon^{S}_3, \epsilon_{\sigma^{\prime}},\epsilon_{\sigma^{\prime\prime}})=(0,0,0,0,0)$. It is important to note that the VEVs of fields $\xi$, $\varphi_{S}$, $\sigma^{\prime}$ and $\sigma^{\prime\prime}$ are invariant under the action of element $S$. In other words, the $A_4$ flavor symmetry is broken down to $Z^{S}_{2}$ by the vacuum of $\xi$, $\varphi_{S}$, $\sigma^{\prime}$ and $\sigma^{\prime\prime}$. Moreover, the alignment direction of $\varphi_S$ is left invariant under the CP transformations $\{X^{0}_{\mathbf{3}},\rho_{\bf 3}(S)X^{0}_{\mathbf{3}}\}$ for $g_{2}g_{3}<0$ and $\{\rho_{\mathbf{3}}(TST^2)X^{0}_{\mathbf{3}}, \rho_{\mathbf{3}}(T^2ST)X^{0}_{\mathbf{3}}\}$ for $g_{2}g_{3}>0$ after the overall phase is extracted. The VEVs $v_{\xi}$, $v_{S}$ and $v_{\sigma}$ are expect to have the same order of magnitude. As we shall show in the following, the correct size of the reactor mixing angle $\theta_{13}$ can be achieved if we choose
\begin{equation}
\label{eq:hierarchy_nu}|v_{\xi}|, |v_S|, |v_{\sigma}|\sim\lambda\Lambda\,.
\end{equation}

\subsection{\label{subsec:model_LO}The model at leading order}

The Yukawa interactions for the charged lepton read,
\begin{eqnarray}
\nonumber&&\hskip-0.35in
w_l=\frac{y_{\tau}}{\Lambda}\tau^ch_{d}(l\varphi_T)_{\mathbf{1}}
+\frac{y_{\mu_1}}{\Lambda^2}\mu^ch_{d}(l(\varphi_T\varphi_T)_{\mathbf{3}_{S}})_{\mathbf{1}}
+\frac{y_{\mu_2}}{\Lambda^2}\mu^ch_{d}(l\varphi_T)_{\mathbf{1}^{\prime}}\zeta^{\prime\prime}
+\frac{y^*_{\mu_2}}{\Lambda^2}\mu^ch_{d}(l\varphi_T)_{\mathbf{1}^{\prime\prime}}\zeta^{\prime}\\
\nonumber&&+\frac{y_{e_1}}{\Lambda^3}e^ch_{d}(l\varphi_T)_{\mathbf{1}}(\varphi_T\varphi_T)_{\mathbf{1}}
+\frac{y_{e_2}}{\Lambda^3}e^ch_{d}(l\varphi_T)_{\mathbf{1}^{\prime}}(\varphi_T\varphi_T)_{\mathbf{1}^{\prime\prime}}
+\frac{y^*_{e_2}}{\Lambda^3}e^ch_{d}(l\varphi_T)_{\mathbf{1}^{\prime\prime}}(\varphi_T\varphi_T)_{\mathbf{1}^{\prime}} \\
\nonumber &&+\frac{y_{e_3}}{\Lambda^3}e^ch_{d}((l\varphi_T)_{\mathbf{3}_S}(\varphi_T\varphi_T)_{\mathbf{3}_S})_{\mathbf{1}}
+\frac{y_{e_4}}{\Lambda^3}e^ch_{d}((l\varphi_T)_{\mathbf{3}_A}(\varphi_T\varphi_T)_{\mathbf{3}_S})_{\mathbf{1}}\\
\nonumber &&+\frac{y_{e_5}}{\Lambda^3}e^ch_{d}(l(\varphi_T\varphi_T)_{\mathbf{3}_{S}})_{\mathbf{1}^{\prime\prime}}\zeta^{\prime}
+\frac{y^*_{e_5}}{\Lambda^3}e^c h_{d}(l(\varphi_T\varphi_T)_{\mathbf{3}_S})_{\mathbf{1}^{\prime}}\zeta^{\prime\prime}  +\frac{y_{e_6}}{\Lambda^3}e^ch_{d}(l\varphi_T)_{\mathbf{1}}\zeta^{\prime}\zeta^{\prime\prime} \\
\label{eq:wl_LO}&&+\frac{y_{e_7}}{\Lambda^3}e^c h_{d}(l\varphi_T)_{\mathbf{1}^{\prime}}\zeta^{\prime2}
+\frac{y^*_{e_7}}{\Lambda^3}e^ch_{d}(l\varphi_T)_{\mathbf{1}^{\prime\prime}}\zeta^{\prime\prime2}+\ldots\,,
\end{eqnarray}
where dots denote the higher dimensional operators which will be
discussed later. The CP symmetry constrains the coupling constants $y_{\tau}$, $y_{\mu1}$, $y_{e1}$, $y_{e3}$ and $y_{e6}$ to be real, $y_{e4}$ to be pure imaginary, and $y_{\mu2}$, $y_{e2}$, $y_{e5}$ and $y_{e7}$ to be general complex numbers. Notice that the auxiliary $Z_4$ symmetry imposes different powers of $\zeta^\prime$, $\zeta^{\prime\prime}$ and $\varphi_T$ for the electron, muon and tau mass terms. Inserting the vacuum configuration of Eq.~\eqref{eq:vev_charged} into the above superpotential $w_{l}$, we find the charged lepton mass matrix is diagonal
\begin{equation}
m_{l}=\left(\begin{array}{ccc}
y_{e}\frac{v^3_{T}}{\Lambda^3} &~ 0 ~& 0 \\
0 &~ y_{\mu}\frac{v^2_{T}}{\Lambda^2} ~& 0 \\
0 &~ 0 ~& y_{\tau}\frac{v_{T}}{\Lambda}
\end{array}\right)v_{d}\,,
\end{equation}
where $v_d=\langle h_d\rangle$ is the VEV of Higgs field $h_{d}$ and parameters $y_{e}$ and $y_{\mu}$ are defined as
\begin{equation}
y_e=y_{e_2}-\frac{2}{9}y_{e_3}+\frac{1}{3}y_{e_4}+\frac{2}{3}y_{e5}\frac{v_{\zeta}}{v_T}+y_{e_7}\frac{v^2_{\zeta}}{v^2_T}, \qquad
y_{\mu}=\frac{2}{3}y_{\mu_1}+y_{\mu_2}\frac{v_{\zeta}}{v_T}\,.
\end{equation}
One can easily see that the realistic mass hierarchy $m_{\tau}:m_{\mu}:m_e\simeq1:\lambda^2:\lambda^4$ is obtained for
$|v_{\zeta}|, |v_T|\sim\mathcal{O}(\lambda^2\Lambda)$. Although the vacuum alignment of $\varphi_T$ breaks the $A_4$ flavor symmetry completely, the hermitian combination $m^{\dagger}_lm_l$ is invariant under the action of $T$, i.e., $\rho_{\mathbf{3}}(T)^{\dagger}m^{\dagger}_{l}m_l\rho_{\mathbf{3}}(T)=m^{\dagger}_{l}m_l$. Hence the $Z^{T}_3$ subgroup is accidently preserved by the charged lepton mass matrix at LO. This accidental symmetry does not survive at the next to the leading order level.

The light neutrino masses are generated by the type-I seesaw mechanism. The most general LO superpotential for the neutrino masses is given by
\begin{equation}\label{eq:neutrino_LO}
 w_{\nu}=y\left(l\nu^c\right)_{\bf{1}}h_u+y_1\left(\nu^c\nu^c\right)_{\mathbf{1}}\xi+y_2\left((\nu^c\nu^c)_{{\bf3}_{S}}\varphi_{S}\right)_{\mathbf{1}}\,,
\end{equation}
where all the couplings are real due to the CP symmetry. With the vacuum alignment of $\xi$ and $\varphi_S$ in Eq.~\eqref{eq:neutrino_vacuum}, the neutrino Dirac and Majorana mass matrices take the form
\begin{equation}
m_{D}=yv_{u}\begin{pmatrix}
1 &~  0 &~ 0 \\
0 &~  0 &~ 1 \\
0 &~  1 &~ 0
\end{pmatrix},\quad m_{M}=\begin{pmatrix}
y_1v_{\xi}+2y_{2}v_{S}/3  &  -y_{2}v_{S}/3  & -y_{2}v_{S}/3 \\
-y_{2}v_{S}/3    & 2y_{2}v_{S}/3   &   y_{1}v_{\xi}-y_{2}v_{S}/3  \\
-y_{2}v_{S}/3   &  y_1v_{\xi}-y_{2}v_{S}/3   &  2y_{2}v_{S}/3
\end{pmatrix}\,,
\end{equation}
where $v_{u}=\langle h_{u}\rangle$. The effective light neutrino mass matrix is given by the see-saw relation
\begin{equation}
m_{\nu}=-m_Dm^{-1}_Mm^{T}_D=U_{TB}\,\text{diag}(m_1,m_2,m_3)U^{T}_{TB}\,,
\end{equation}
with
\begin{equation}\label{eq:neutrino_mass_LO}
m_1=-\frac{y^2v^2_u}{y_1v_{\xi}+y_2v_S},\quad
m_2=-\frac{y^2v^2_u}{y_1v_{\xi}},\quad
m_3=\frac{y^2v^2_{u}}{y_1v_{\xi}-y_2v_S}\,.
\end{equation}
We see that the above neutrino masses fulfill the sum rule
\begin{equation}
\label{eq:sumrule}
\frac{1}{m_1}-\frac{1}{m_3}=\frac{2}{m_2}\,.
\end{equation}
In the case of $g_2g_3>0$, the phase difference of $v_{S}$ and $v_{\xi}$ is $\pm\pi/2$, such that the neutrino masses would be partially degenerate with $|m_1|=|m_3|$. Hence we shall be concerned with the scenario of $g_2g_3<0$ in the following. Thus the VEVs $v_{S}$ and $v_{\xi}$ carry the same phase up to $\pi$, and they can be considered as real. The two squared mass gaps $\delta m^2$ and $\Delta m^2$ can be written as
\begin{eqnarray}
\nonumber&&\delta m^2\equiv|m_2|^2-|m_1|^2=\left|\frac{y^2v^2_u}{y_1v_{\xi}}\right|^2\frac{x^2+2x}{\left(1+x\right)^2}\,,\\
&&\Delta m^2\equiv|m_3|^2-\frac{1}{2}(|m_1|^2+|m_2|^2)=\left|\frac{y^2v^2_u}{y_1v_{\xi}}\right|^2\frac{x(6+3x-x^3)}{2\left(1-x^2\right)^2}\,,
\end{eqnarray}
where the parameter $x\equiv y_{2}v_S/(y_1v_{\xi})$ is real. Since the charged lepton mass matrix is diagonal, the effective Majorana mass $m_{ee}$ is exactly the absolute value of the (11) entry of $m_{\nu}$, i.e.
\begin{equation}
m_{ee}=\left|(m_{\nu})_{11}\right|=\left|\frac{y^2v^2_u}{y_1v_{\xi}}\right|\left|\frac{3+x}{3\left(1+x\right)}\right|.
\end{equation}
Using the experimental best fit value $\delta m^2=7.54\times10^{-5}\text{eV}^2$ and $\Delta m^2=2.43(-2.38)\times10^{-3}\text{eV}^2$ for NO (IO) spectrum~\cite{Capozzi:2013csa}, we find three solutions for $x$,
\begin{equation}
x\simeq0.787\,, 1.199\,, -2.014\,,
\end{equation}
where the first two solutions correspond to NO neutrino mass spectrum, while the last one is for the IO spectrum. The resulting predictions for the light neutrino masses and the effective mass $m_{ee}$ are listed in table~\ref{tab:LO_predictions}. Since the leading order PMNS mixing matrix is the tri-bimaximal pattern which gives rise to a vanishing $\theta_{13}$, the Dirac phase can not be fixed uniquely and moderate corrections to $\theta_{13}$ are necessary in order to be in accordance with experimental data.

\begin{table} [t!]
\begin{center}
\begin{tabular}{|c|c|c|c|c|c|}
\hline\hline
$x$ &  $|m_1|\text{(meV)}$ &
 $|m_2|\text{(meV)}$ & $|m_3|\text{(meV)}$ & $m_{ee}\text{(meV)}$
 &
   \text{mass ordering} \\ \hline
0.787  &  5.865 & 10.478 & 49.113  &  7.403 & \text{NO} \\ \hline
1.199 &  4.433 & 9.750 & 48.963  &  6.055 & \text{NO} \\ \hline
$-2.014$ &  51.612 & 52.338 & 17.365 & 16.962 & \text{IO}  \\
 \hline \hline
\end{tabular}
\caption{\label{tab:LO_predictions}The LO predictions for the light neutrino masses $|m_i| (i=1,2,3)$ and the effective Majorana mass
$m_{ee}$ in $0\nu\beta\beta$ decay.}
\end{center}
\end{table}

\subsection{\label{subsec:model_NLO}Subleading order corrections}

The above LO superpotentials $w^{l}_{d}$, $w^{\nu}_{d}$, $w_{l}$ and $w_{\nu}$ receive corrections from higher dimensional operators,  compatible with all the symmetries of the model, which are suppressed by additional powers of the cut-off $\Lambda$. The NLO corrections to the driving superpotentials $w^{l}_{d}$ and $w^{\nu}_{d}$ induce deviations from the LO alignment configuration. Taking into account that the VEVs of the flavon fields in the neutrino and charged lepton sectors are of order $\lambda\Lambda$ and $\lambda^2\Lambda$ respectively, we find that the subleading corrections to $w^{\nu}_{d}$ arise from the unique operator,
\begin{equation}\label{eq:NLO_wd_nu}
\delta w^{\nu}_d=\frac{r}{\Lambda}\sigma^{0}\xi\sigma^{\prime}\sigma^{\prime\prime}\,.
\end{equation}
As we impose the CP symmetry on the theory in the unbroken phase, the coupling $r$ is a real number. The new minimum for $\xi$, $\varphi_S$, $\sigma^{\prime}$ and $\sigma^{\prime\prime}$ is obtained by searching for the zeros of the $F-$terms, the first derivatives of $w^{\nu}_d+\delta w^{\nu}_d$ associated to the driving fields $\xi^0$, $\sigma^0$ and $\varphi^0_S$. We look for a solution which perturbes the LO vacuum in Eq.~\eqref{eq:neutrino_vacuum} to the first order in the $1/\Lambda$ expansion,
\begin{equation}
\langle\xi\rangle=v_{\xi},\quad  \langle\varphi_S\rangle=\left(\begin{array}{c}
v_S+\delta v_{S} \\
v_{S}+\delta v_{S} \\
v_S+\delta v_{S}
\end{array}\right),\quad \langle\sigma^{\prime}\rangle=v_{\sigma}+\delta v_{\sigma^{\prime}},\quad
\langle\sigma^{\prime\prime}\rangle=-g_{4}v_{\sigma}/g_{5}+\delta v_{\sigma''}\,,
\end{equation}
where $v_{\xi}$ is undetermined with
\begin{equation}\label{eq:vev_neutrino_NLO}
\delta v_{S}=-\frac{rM}{2g_{1}g_{2} \Lambda}v_S, \qquad
 \delta v_{\sigma^{\prime}}=\delta v_{\sigma^{\prime\prime}}=0\,.
\end{equation}
Obviously the shift $\delta v_{S}$ carries the same phase as $v_{S}$, thus
the correction to $\langle\varphi_S\rangle$ is proportional the LO VEV and can be absorbed in a redefinition of the parameters $g_2$ and $g_3$. Similarly at the next order the higher dimensional operators contributing to $w^{l}_{d}$ comprise five flavons, two of them belong to the set $\{\zeta^{\prime},\zeta^{\prime\prime},\varphi_{T}\}$ in order to saturate the $Z_4$ charge and two flavons from the set $\{\varphi_S,\xi\}$ together with one field of the type $\{\sigma^{\prime},\sigma^{\prime\prime}\}$, e.g.
\begin{equation}
\zeta^{0}(\varphi^2_{l}\varphi^2_\nu\varphi^{\prime}_{\nu})_{\mathbf{1}}/\Lambda^3,\qquad
(\varphi^{0}_{T}\varphi^2_{l}\varphi^2_\nu\varphi^{\prime}_{\nu})_{\mathbf{1}}/\Lambda^3\,,
\end{equation}
where $\varphi_{l}=\{\zeta^{\prime},\zeta^{\prime\prime},\varphi_{T}\}$, $\varphi_{\nu}=\{\varphi_S,\xi\}$ and $\varphi^{\prime}_{\nu}=\{\sigma^{\prime},\sigma^{\prime\prime}\}$, and all the possible combinations and possible $A_4$ contractions should be considered. As a result, the subleading contributions to the $F-$terms of the driving fields $\zeta^0$ and $\varphi^0_T$ are suppressed by $\langle\varphi_{\nu}\rangle^2\langle\varphi^{\prime}_{\nu}\rangle/\Lambda^3\sim\lambda^3$ with respect to the contributions from the LO terms in Eq.~\eqref{eq:driving_pote_clepton}. Therefore the vacuum of $\varphi_T$ and $\zeta^{\prime\prime}$ acquire corrections of relative order $\lambda^3$ and the shifted vacuum can be parameterized as
\begin{equation}
\label{eq:vacuum_clep_NLO}\langle\varphi_T\rangle=\left(\begin{array}{c}
\alpha_1\lambda^3 \\
1+\alpha_2\lambda^3 \\
\alpha_3\lambda^3
\end{array}\right)v_T\,,
\qquad \langle\zeta^{\prime\prime}\rangle=\alpha_4\lambda^3v_{\zeta}\,,
\end{equation}
where $\alpha_i (i=1,2,3,4)$ are generally complex numbers with absolute value of order one.

As regards the corrections to $w_l$, the mass terms related to $\tau^{c}$, $\mu^{c}$  and $e^{c}$ require one, two and three flavon fields from the set $\{\zeta^{\prime},\zeta^{\prime\prime},\varphi_{T}\}$ respectively in order to fulfill the invariance under the $Z_4$ component of the
flavor symmetry group, and the higher dimensional operators can be obtained by further multiplying the combination $\varphi^2_{\nu}\varphi'_{\nu}$ in all possible ways. Therefore the subleading operators contributing to the charged lepton masses take the form
\begin{equation}
\delta w_l=\tau^ch_{d}(l\varphi_l\varphi^2_{\nu}\varphi^{\prime}_{\nu})_{\mathbf{1}}/\Lambda^4+\mu^ch_{d}(l\varphi^2_{l}\varphi^2_{\nu}\varphi^{\prime}_{\nu})_{\mathbf{1}}/\Lambda^5+
e^ch_{d}(l\varphi^3_{l}\varphi^2_{\nu}\varphi^{\prime}_{\nu})_{\mathbf{1}}/\Lambda^6\,,
\end{equation}
where all dimensionless coupling constants are omitted. The charged lepton mass matrix is obtained by plugging the LO vacuum to these new operators plus the contribution of the LO superpotential in Eq.~\eqref{eq:wl_LO}, evaluated with the shifted VEVs of Eq.~\eqref{eq:vacuum_clep_NLO}. Thus we find the corrected charged lepton mass matrix has the following structure
\begin{equation}\label{eq:ml_NLO}
m_{l}=\begin{pmatrix}
\mathcal{O}(\lambda^6)  &~  \mathcal{O}(\lambda^9)  &~  \mathcal{O}(\lambda^9)  \\
\mathcal{O}(\lambda^7)  &~  \mathcal{O}(\lambda^4)  &~  \mathcal{O}(\lambda^7) \\
\mathcal{O}(\lambda^5)  &~ \mathcal{O}(\lambda^5)  &~ \mathcal{O}(\lambda^2)
\end{pmatrix}v_d\,,
\end{equation}
where only the order of magnitude of each entry is presented. As a result, the unitary matrix $U_{l}$ which realizes the transformation to the physical basis where the mass matrix $m^{\dagger}_lm_l$ is diagonal at NLO is of the general form
\begin{equation}
U_l\simeq\begin{pmatrix}
1 &~ V_{12}\lambda^3 &~ V_{13}\lambda^3 \\
-V^{*}_{12}\lambda^3 &~ 1 &~ V_{23}\lambda^3 \\
-V^{*}_{13}\lambda^3 &~ -V^{*}_{23}\lambda^3 &~ 1
\end{pmatrix}\,,
\end{equation}
where $V_{ij}$ are unspecified order one constant. Therefore the contributions of the charged lepton sector to the lepton mixing angles is of order $\lambda^3$ and can be safely neglected.

In the same manner we can analyze the subleading corrections in the neutrino sector. The NLO operators contributing to the neutrino Dirac mass is given by
\begin{equation}
\left(l\nu^{c}\varphi^2_{\nu}\varphi^{\prime}_{\nu}\right)_{\mathbf{1}}h_u/\Lambda^3\,,
\end{equation}
where all possible independent $A_4$ contractions should be considered. The resulting contributions are suppressed by $\lambda^3$ compared to the LO term $\left(l\nu^c\right)_{\mathbf{1}}h_u$ in Eq.~\eqref{eq:neutrino_LO} and consequently are negligible. The NLO corrections to the right-handed Majorana neutrino masses are
\begin{equation}\label{eq:wnu_NLO}
\delta w_{\nu}=\frac{y_3}{\Lambda}(\nu^c\nu^c)_{\mathbf{1}}\sigma^{\prime}\sigma^{\prime\prime}
+\frac{y_4}{\Lambda}(\nu^c\nu^c)_{\mathbf{1}^{\prime}}(\sigma^{\prime})^2+\frac{y^*_4}{\Lambda}(\nu^c\nu^c)_{\mathbf{1}^{\prime\prime}}(\sigma^{\prime\prime})^2\,,
\end{equation}
where the coupling $y_{3}$ is real and $y_{4}$ is complex because of the invariance under the CP symmetry. Since the structure of the LO vacuum of the neutrino flavons is unchanged by the NLO corrections, the only possible modification to the neutrino masses arises from the operators listed in Eq.~\eqref{eq:wnu_NLO}. Inserting the alignment of $\sigma^{\prime}$ and $\sigma^{\prime\prime}$ into these higher dimensional operators and taking into account the LO contribution, we find that the neutrino Majorana mass matrix becomes
\begin{equation}
 m_M=\begin{pmatrix}
 a+2 b &~ c-b ~& d-b \\
 c-b &~ 2 b+d ~& a-b \\
 d-b &~ a-b ~& 2 b+c \\
\end{pmatrix}\,.
\end{equation}
with
\begin{equation}
\begin{split}
&a=y_{1}v_{\xi}-\frac{y_3g_4  v^2_{\sigma} }{g_5 \Lambda }=(y_{1}-\frac{y_{3}M}{g_{1}\Lambda})v_{\xi}, \qquad b=\frac{y_{2}v_{S}}{3},\\
& c=\frac{y_4 v_{\sigma}^2 }{\Lambda}=\frac{y_4g^{*}_4Mv_{\xi}}{g_1 g_4 \Lambda }, \qquad d=\frac{y^*_4g^2_{4} v_{\sigma}^2 }{g^2_{5}\Lambda}=\frac{y^*_4g_4 M v_{\xi}}{g_1g^{*}_4 \Lambda}\,.
\end{split}
\end{equation}
We see that the common phase of $v_{S}$ and $v_{\xi}$ is an overall phase of $m_{M}$, and consequently it is irrelevant for neutrino masses and the lepton flavor mixing. Thus the parameters $a$ and $b$ can be considered as real and $c$ and $d$ are complex with $d=c^{*}$ after the overall phase is factored out. Moreover, as $c$ and $d$ originate from the NLO operators in Eq.~\eqref{eq:wnu_NLO}, they are suppressed by $\lambda$ with respect to the LO contributions $a$ and $b$, i.e.
\begin{equation}
 |a|,~|b|\sim \lambda\Lambda, \qquad |c|,~|d|\sim \lambda^{2}\Lambda\,.
\end{equation}
Finally we obtain that the light neutrino mass matrix $m_{\nu}$ including the subleading corrections is modified into
\begin{equation}
\label{eq:mnu_NLO}m_{\nu}=\alpha\begin{pmatrix}
2  &~  -1  &~  -1  \\
-1 &~  2   &~  -1  \\
-1 &~   -1 &~ 2
\end{pmatrix}+\beta\begin{pmatrix}
1 &~  0  &~  0  \\
0 &~  0  &~  1  \\
0 &~  1  &~ 0
\end{pmatrix}+\gamma\begin{pmatrix}
0 &~ 1  &~ 1  \\
1 &~ 1  &~ 0  \\
1 &~ 0  &~ 1
\end{pmatrix}+\delta\begin{pmatrix}
0 &~  1 &~  -1  \\
1 &~  -1 &~ 0   \\
-1&~  0 &~  1
\end{pmatrix}\,.
\end{equation}
The parameters $\alpha$, $\beta$, $\gamma$ and $\delta$ are expressed in terms of $a$, $b$, $c$, $d$ as
\begin{equation}\label{eq:parameters}
\begin{split}
\alpha&=\frac{b }{a^2-a (c+d)-9 b^2+c^2-c d+d^2}\,,\\
\beta&=\frac{ -a^2+3 b^2+c d}{(a+c+d) \left[a^2-a (c+d)-9 b^2+c^2-c d+d^2\right]}\,, \\
\gamma&=\frac{ a (c+d)+6 b^2-c^2-d^2}{2 (a+c+d) \left[a^2-a (c+d)-9 b^2+c^2-c d+d^2\right]}\,, \\
\delta&=\frac{ c-d}{2 \left[a^2-a (c+d)-9 b^2+c^2-c d+d^2\right]}\,,
\end{split}
\end{equation}
where an overall factor $y^2v^2_u$ has been omitted. One sees that $\alpha$, $\beta$ and $\gamma$ are real while $\delta$ is pure imaginary up to an overall phase of $m_{\nu}$. It is remarkable that this neutrino mass matrix has the most general form of case I in which the residual symmetry of the neutrino sector is $Z^{S}_2\times X_{\nu}$ with $X_{\nu}=X^{0}_{\mathbf{3}}$ or $\rho_{\mathbf{3}}(S)X^{0}_{\mathbf{3}}$. Hence the results for mixing angles are given by Eq.~\eqref{eq:mixing_angles}, the Dirac CP violation phase and the atmospheric mixing angle are maximal while both Majorana CP phases are trivial. The reactor mixing angle $\theta_{13}$ depends on the angle $\theta$ which is determined to be
\begin{eqnarray}
\tan2\theta=\frac{i\delta}{\sqrt{3}\alpha}=\frac{i(c-d)}{2\sqrt{3}b}\sim \mathcal{O}(\lambda)\,,
\end{eqnarray}
in our model. As a consequence, the correct order of $\theta_{13}$ is naturally achieved.

\section{\label{sec:conclusion}Conclusion}

The $A_4$ group has been widely used to study the lepton mixing. In the present work we discuss the interplay between $A_4$ and CP symmetry. Generally two CP symmetries given by Eqs.~(\ref{eq:CP_first}, \ref{eq:CP_second}) can be consistently combined with the $A_4$ flavor group. The first one was considered in previous literature~\cite{Ding:2013bpa}, and it can be imposed on a generic $A_4$ model regardless of the matter content. The second possible CP symmetry interchanges the $A_4$ representations $\mathbf{1}'$ and $\mathbf{1}''$. As a consequence, if one intends to impose $A_4$ as well as this CP symmetry, both fields $\varphi_{\mathbf{1}'}$ and $\varphi_{\mathbf{1}''}$ should be present or absent simultaneously and they should carry the same quantum numbers under all symmetries of the model except $A_4$.

We have analyzed the lepton mixing patterns that arise from a theory in which $A_4$ and the second compatible CP symmetry are broken to residual groups $Z_3$ and $Z_2\times CP$ in the charged lepton and neutrino sectors, respectively. It is sufficient to only consider the remnant symmetries $G_{l}=Z^{T}_{3}$ and $G_{\nu}=Z^{S}_2\times X_{\nu}$ with $X_{\nu}=X^{0}_{\mathbf{3}}$,~$\rho_{\mathbf{3}}(S)X^{0}_{\mathbf{3}}$,~$\rho_{\mathbf{3}}(TST^2)X^{0}_{\mathbf{3}}$,~$\rho_{\mathbf{3}}(T^2ST)X^{0}_{\mathbf{3}}$, since other possible choices of $G_{l}$ and $G_{\nu}$ are related by similarity transformations to this representative one and thus don't lead to new results. We find one interesting mixing pattern which can accommodate the experimental data on lepton mixing angles for certain values of the parameter $\theta$. In this case, the atmospheric mixing angle $\theta_{23}$ and Dirac CP phase $\delta_{CP}$ are predicted to be maximal, both Majorana phases are trivial, and the solar and reactor mixing angles satisfy the sum rule $\sin^2\theta_{12}\cos^2\theta_{13}=1/3$.

Furthermore we construct a see-saw model for lepton based on the flavor symmetry $A_4$ and the second possible CP symmetry compatible with $A_4$. The lepton mixing matrix is exactly the tri-maximal mixing at leading order. The tau, muon and electron masses originate from operators with one, two and three flavons respectively because of the auxiliary symmetry $Z_3\times Z_5$, such that the observed mass hierarchy among the charged leptons is achieved. Subleading contributions give rise to a non-vanishing reactor mixing angle $\theta_{13}$ which is predicted to be of the correct order $\lambda$ in our model. The interesting mixing pattern found in model independent analysis is naturally reproduced after the higher order corrections induced by higher dimensional operators are considered. Finally it is interesting to consider other residual symmetries distinct from $(G_{l}, G_{\nu})=(Z^{T}_3, Z^{S}_2\times X_{\nu})$ and the application in model building such that new predictions for the CP phases could be obtained.

\section*{Acknowledgements}

We are grateful to Jose W.F. Valle for enlightening discussions on Babu-Ma-Valle model which motivates this work. One of the author (G.J.D.) acknowledges Stephen F. King and Alexander J. Stuart for collaboration on generalized CP symmetry and $A_4$. This work is supported by the National Natural Science Foundation of China under Grant Nos. 11275188, 11179007 and 11522546.

\end{document}